\documentstyle[12pt,epsf]{article}
\begin{document}
\renewcommand{\vec}[1]{{\bf #1}}
\newlength{\figwidth}
\newlength{\figheight}
\setlength{\figwidth}{0.5 \textwidth}
\addtolength{\figwidth}{-0.5 \columnsep}
\setlength{\figheight}{ 0.9 \figwidth}
\pagestyle{empty}

\begin{flushleft}
\large \bf
Non-degenerate Two Photon Absorption Spectra \\
of Si Nanocrystallites
\end{flushleft}

\begin{flushleft}
Toshiaki Iitaka and Toshikazu Ebisuzaki
\end{flushleft}

\begin{flushleft}
Computational Science Laboratory \\
The Institute of Physical and Chemical Research (RIKEN) \\
2-1 Hirosawa, Wako, Saitama 351-0198, JAPAN \\
e-mail: tiitaka@postman.riken.go.jp, \ \ URL: http://espero.riken.go.jp/
\end{flushleft}



We propose an efficient linear-scaling time-dependent method for calculating nonlinear response function, and study the size effects in non-degenerate two photon absorption spectra of Si nanocrystallites by using semi-empirical pseudopotentials.  


\begin{flushleft}
\bf
1. INTRODUCTION
\end{flushleft}


The  two-photon absorption (TPA) spectrum is an important optical property of semiconductors. It brings valuable information complementary to the one-photon absorption spectrum, gives the limits to the optical transparency of materials, and causes laser-induced damages.

Recently, the TPA spectrum of direct-gap nanocrystallites was calculated by Cotter et al. \cite{Cotter} by using the effective-mass approximation to predict a strong effect of the nanocrystallites size.  However the experimental result \cite{Banfi} does not support this result. Since then TPA spectra of nanocrystallites has been attracting attention of researchers.

In this paper, we develop a new algorithm for calculating TPA spectra
by using semi-empirical local pseudopotentials \cite{Wang94a}, finite
difference method in real space \cite{Cheli94}, and a linear-scaling time-dependent method which has been applied to the calculation of the
linear-response functions \cite{Iitaka95,Tanaka98,Iitaka97,Nomura97}.
This efficient algorithm made it possible, for the first time,
to calculate the size effect on the TPA spectra of very large
nanocrystallites without using effective-mass approximation.
In the following we show the effectiveness of this algorithm by
applying it to the TPA spectra of indirect-gap nanocrystallites.
Though this result does not solve the controversy related to direct-gap nanocrystallites \cite{Cotter,Banfi},
it will be an important step toward it.

\begin{flushleft}
\bf
2. TPA COEFFICIENT
\end{flushleft}

The non-degenerate TPA coefficient $\beta_{12}(\omega_1,\omega_2)$ describes absorption of the probe light with the frequency $\omega_1$ and the polarization $\vec{e}_a$ in the presence of the excite light with the frequency $\omega_2$ and the polarization $\vec{e}_b$, and, in the transparent region ($\omega_1,\omega_2<E_g$), it is related to the third order nonlinear susceptibility $ \chi^{(3)}_{abba}$ \cite{Cotter90} by
\begin{eqnarray}
\beta_{ab}(\omega_1,\omega_2) 
&=& \frac{6(2\pi)^2\omega_1}{c^2 \eta_{aa}(\omega_1) \eta_{bb}(\omega_2)}
{\rm Im} \chi^{(3)}_{abba}(-\omega_1;-\omega_2,\omega_2,\omega_1)
\end{eqnarray}
where $\eta_{aa}(\omega_1)$ and $\eta_{bb}(\omega_2)$ are the real part of linear refractive index.
A simple form of $\chi^{(3)}$ for TPA is provided by the second-order time-dependent perturbation theory \cite{Cotter90},
\begin{eqnarray}
\label{eq:chi3}
\chi^{(3)}_{abba}(-\omega_1;-\omega_2,\omega_2,\omega_1)
&=&
\sum_{v,c} \frac{1}{6 \hbar \epsilon_0 V}
\frac{\alpha_{vc}(\omega_1,\omega_2)\alpha_{vc}^*(\omega_1,\omega_2)}{\omega_{cv}-i\gamma-\omega_1-\omega_2}
\end{eqnarray}
where the composite matrix element is defined by
\begin{eqnarray}
\alpha_{vc}(\omega_1,\omega_2)
&=&
\frac{e^2}{\hbar} \sum_m \left[
\frac{(\vec{e}_a\cdot\vec{r}_{cm})(\vec{e}_b\cdot\vec{r}_{mv})
}{\omega_{mv}-\omega_2}
+
\frac{(\vec{e}_b\cdot\vec{r}_{cm})(\vec{e}_a\cdot\vec{r}_{mv})
}{\omega_{mv}-\omega_1}
\right]
.
\end{eqnarray}
The subscripts $v$ and $c$ in the summation run over all valence-band states and all conduction-band states, respectively.
The subscript $m$ for the intermediate states runs over both valence- and conduction-band states.

It is worth to note that the form  Eq. (\ref{eq:chi3}) looks very similar to that of the linear-response function, and that, therefore, we may calculate $\chi^{(3)}$ by using the linear-scaling time-dependent method for the linear-response functions \cite{Iitaka97,Nomura97},
\begin{equation}
\label{eq:chi.time.one}
\label{eq:chi.numerical.1}
\hspace*{-5mm}
\chi^{(1)}_{ba}(\omega) 
 \!\! = \!\! 
\sum_{v,c} \frac{-e^2}{\hbar \epsilon_0 V}
\left\{
\frac{(\vec{e}_b\cdot\vec{r}_{vc})(\vec{e}_a\cdot\vec{r}_{cv})}{\omega_{cv}-i\gamma-\omega}  - 
\frac{(\vec{e}_b\cdot\vec{r}_{vc})(\vec{e}_a\cdot\vec{r}_{cv})}{\omega_{cv}+i\gamma+\omega} 
\right\}
 \!\! = \!\! 
\left\langle \!\!\! \left\langle
\rule{0pt}{24pt}
\int_0^T \!\!\!dt \ \ e^{+i(\omega + i\gamma)t}  \delta B(t)
\right\rangle \!\!\! \right\rangle
\end{equation}
where $\left\langle \left\langle \   \cdot \  \right\rangle \right\rangle \ $ indicates the statistical average over random vectors, and $\delta B(t)$ is the response of the system defined by
\begin{equation}
\label{eq:correlation}
\delta B(t)  = 2 \ {\rm Im}
\langle \Phi_{E_f} | e^{+iHt} (\vec{e}_b\cdot\vec{r}) 
e^{-iHt} \theta(H-E_f) (\vec{e}_a \cdot\vec{r})  | \Phi_{E_f} \rangle 
.
\end{equation}
The ket $| \Phi_{E_f} \rangle $ is a random vector projected onto the Fermi occupied states, and $\theta(H-E_f)$ is a projection operator to extract the Fermi unoccupied states. The imaginary part of frequency $\gamma$ is introduced to limit the integration time in Eq. (\ref{eq:chi.numerical.1}) to a finite value $T=-\ln \delta / \gamma $ with $\delta$ being the relative numerical accuracy we need.
The main difference of $\chi^{(3)}$ from $\chi^{(1)}$ is the composite matrix elements in place of the dipole matrix elements, which we can calculate with the help of the particle source method \cite{Iitaka95,Tanaka98}.

\begin{flushleft}
\bf
3. RESULT
\end{flushleft}
\noindent
\ \vspace*{-\figheight}
\begin{minipage}[t]{\figwidth}
\hspace*{\parindent}
Figure~1 shows the non-degenerate TPA coefficient $\beta_{xx}(\omega_1,\omega_2)$ of hydrogenated cubic Si nanocrystallites of size $l=1 \sim 4$ (nm) as a function of $\omega_1$ with a fixed excite light frequency, $\omega_2 = 2.4 \mbox{(eV)}$.  
In the calculation, we used the Hamiltonian matrix discretized into $N=L^3$ ($L=32\sim80$) cubic meshes in real space, which consists of the semi-empirical local pseudopotential \cite{Wang94a} the kinetic energy operator in the finite difference form \cite{Cheli94} .
The results were averaged over $2-16$ random vectors depending on the system size.
The energy resolution is set to $\gamma=200 \mbox{(m eV)}$, which may not small enough to resolve the fine structures in the spectra but small enough to study the size effects on the magnitude of ${\rm Im } \chi^{(3)}$
\end{minipage} 
\hspace{\columnsep}%
%
%
\begin{minipage}[t]{\figwidth}
\vspace*{0.01\figheight}
\epsfile{file=nondeg.eps,height=\figheight}
\begin{flushleft}
{\bf Fig. 1} \ $\beta_{xx}(\omega_1,\omega_2)$ of Si nanocrystallites with $\omega_2 = 2.4 \mbox{(eV)}$.
\end{flushleft}
\end{minipage}
\clearpage
The size effects on the TPA coefficient is evident in the figure.
The absorption increases as the crystallite size increases, and approaches to the bulk value when $ l \approx 4 $ (nm). 
The tail extending below the TPA absorption edge $0.8 \mbox{(eV)} $ in the spectrum is due to the Lorentzian distribution with the finite width $\gamma$ in the time-dependent calculation.

\begin{flushleft}
\bf
4. DISCUSSIONS AND SUMMARY
\end{flushleft}
The computational cost of our method is $O(MN)$ for the linear response function and the non-degenerate TPA coefficient and $O(M^2N)$ for the degenerate TPA coefficient. The large number $M \propto T/\Delta t \propto E_{max}/\gamma \ll N$ is the number of time steps in a time evolution or the number of the frequency $\omega_1$ to be calculated, where $E_{max}$ is the range of the eigenenergy.
Therefore our method is much more efficient than the $O(N^3)$ diagonalization method \cite{Misao94} and the {\it equation of motion methods} \cite{Hobbs96} whose computational effort is $O(M^2N)$ for the linear-response function and $O(M^3N)$ for the third order susceptibility.

In summary, we have established an efficient linear-scaling
time-dependent method for nonlinear response function, and studied the
size effects on the two photon absorption spectra of Si
nanocrystallites. Such a large scale calculation has been impossible
with conventional algorithms.
Therefore the present result will be an important step to solve the controversy related to direct-gap nanocrystallites \cite{Cotter,Banfi}.

The calculation in this article has been done on the supercomputer Fujitsu VPP500 at RIKEN and NIG.



\end{document}